**Does Artificial Intelligence benefit UK businesses? An empirical study of the impact of AI on productivity.**

Sam Hainsworth

17/02/2023

## Abstract

Media hype and technological breakthroughs are fuelling the race to adopt Artificial Intelligence amongst the business community, but is there evidence to suggest this will increase productivity? This paper uses 2015-2019 microdata from the UK Office for National Statistics to identify if the adoption of Artificial Intelligence techniques increases labour productivity in UK businesses. Using fixed effects estimation (Within Group) with a log-linear regression specification the paper concludes that there is no statistically significant impact of AI adoption on labour productivity.

## Section 1: Introduction

The Artificial Intelligence (AI) market is growing rapidly, with annual investment in UK AI companies increasing from £252 million in 2015 to over £2.8 billion in 2021[1]. On a global scale, it is estimated AI funding has reached $66.8 billion in 2021 - double 2020 figures[2].

Alongside commercial growth, there has been a substantial increase in AI research activity and discovery. Shown in Figure 1, the number of peer-reviewed AI publications has increased from less than 40,000 in 2010 to over 120,000 in 2019[3]. AI is a General Purpose Technology[4], with applications across many scientific and commercial areas.

*Figure 1: Number of peer-reviewed AI publications worldwide, 2000-2019*

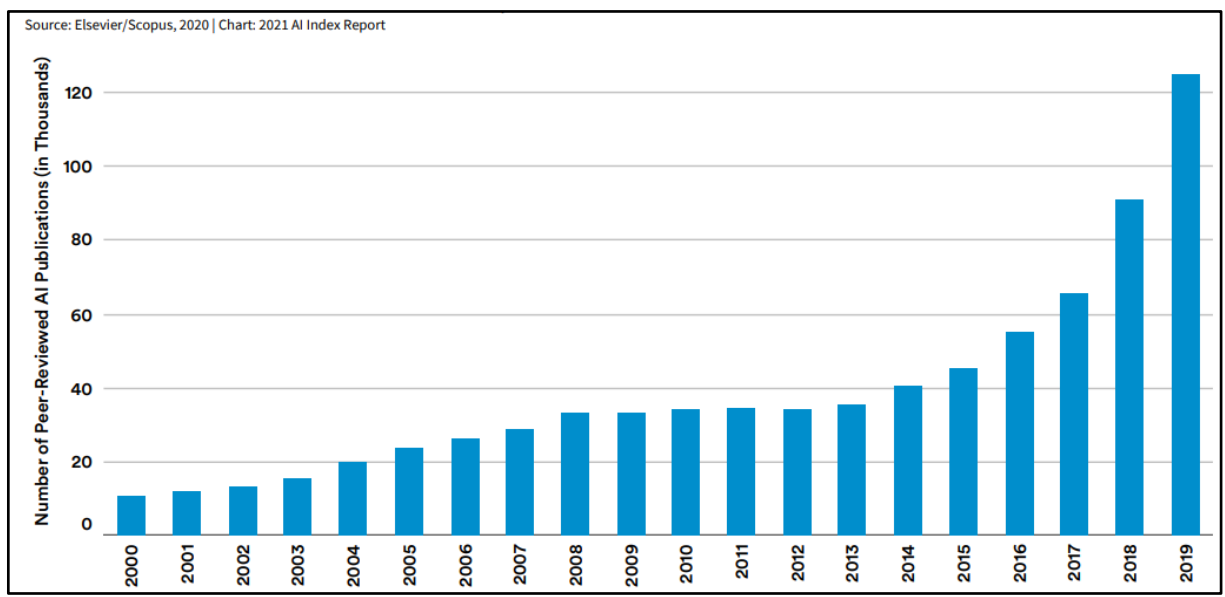

This rapid growth, and the opportunities that AI offers, has come to the attention of policymakers' worldwide. The UK's Office for AI (OAI) was established in 2018 to 'drive responsible, innovative uptake of AI technologies for the benefit of everyone in the UK'[5]. Officials are responsible for developing government policy that will stimulate the innovation and adoption of AI technologies, enabling all regions and sectors of the UK to benefit.

Robust and insightful analysis can support the OAI in many decisions, including the allocation of spending, policy design and development, and monitoring and evaluation of existing policies.

## Section 2: Aims, objectives and project scope

This research will address two analytical questions that are fundamental to the OAI achieving its stated objectives:

1. What are the key characteristics of firms adopting AI?

---

[1] Beauhurst data, DCMS analysis
[2] Top Global Artificial Intelligence Markets, US Department for Commerce (2022)
[3] Artificial Intelligence Index Report 2021, Human-Centered AI - Stanford University (2021)
[4] General Purpose Technologies - Handbook of Economic Growth, Jovanovic, Boyan, Rousseau (2005)
[5] About Us - Office for Artificial Intelligence, DCMS - Link

2. What is the impact of adopting AI technologies on labour productivity?

Using economic theory, combined with robust econometric analysis, this research will provide evidence to answer these research questions and discuss the implications for policy. This research will aim to provide new insights into the decision making of firm's using AI and quantitative evidence of the effects AI is having on UK businesses. To effectively analyse these questions, there are four objectives that this paper will aim to achieve:

1. Develop theoretical framework for AI adoption and business decision making
2. Create a unique panel micro-dataset for UK businesses, including AI adoption and key business variables
3. Establish and estimate an econometric specification to answer the research questions
4. Interpret and consider the key policy implications of the findings

The paper will assess relevant economic literature to develop a theoretical underpinning for the analysis. From the macroeconomic perspective this will look at technology as a driver of economic growth, and from the microeconomic perspective it will focus on firm-level evidence of the impact of AI technologies on business outcomes. A summary of the data and research methodology will follow. The findings will be discussed, interpreted and the implications for policy will be highlighted. Data limitations, further analysis and recommendations for next steps will also be provided alongside conclusions from the study.

## Section 3: Literature review or state of the prior art

**The Importance of technology for productivity and growth**

Economic growth is a key objective for many governments as a means of raising living standards. Economists have long studied the determinants of growth and successive theories have developed a consensus that the main determinant of long-term growth is technological progress.

Solow's growth model demonstrates labour-augmenting technology as the only determinant of growth in long-run per-capita output[6]. Building on this, Romer's endogenous growth model highlights the importance of technology and innovation in driving long-run economic growth and focusses importance on the decision of allocating labour resources to R&D activities[7] [8].

These theoretical findings show that the recent wave of innovation in digital technologies has the potential to increase productivity and contribute to economic growth. However, much of the emerging real-world evidence is yet to validate this hypothesis. Coyle and Mei identify the Information and Communication industry, and specifically the software sub-sector, as being one of two main industries to have contributed to the productivity slowdown since 2008[9]. Aggregate UK productivity measures also show that there may be a more complex relationship between digital technology innovation and the impact on business productivity.

---

[6] Technical progress and Growth, R. Solow (1957)
[7] Endogenous Technological Change, P. Romer (1990)
[8] Technology and economic growth: From Robert Solow to Paul Romer, Zhao (2018)

Analysis by the Bank of England has shown that firm-level productivity is not uniformly distributed and that there are a small number of frontier firms with very high productivity growth and large long tail of firms with little or no productivity growth since 2000[10]. These findings are supported by research on the heterogeneity of firm-level performance and the continued divergence of firms' productivity[11]. Figure 2 shows a time series chart, highlighting this productivity divergence of frontier firms (shown in orange) versus the rest, since 2002.

*Figure 2: UK business productivity over time*

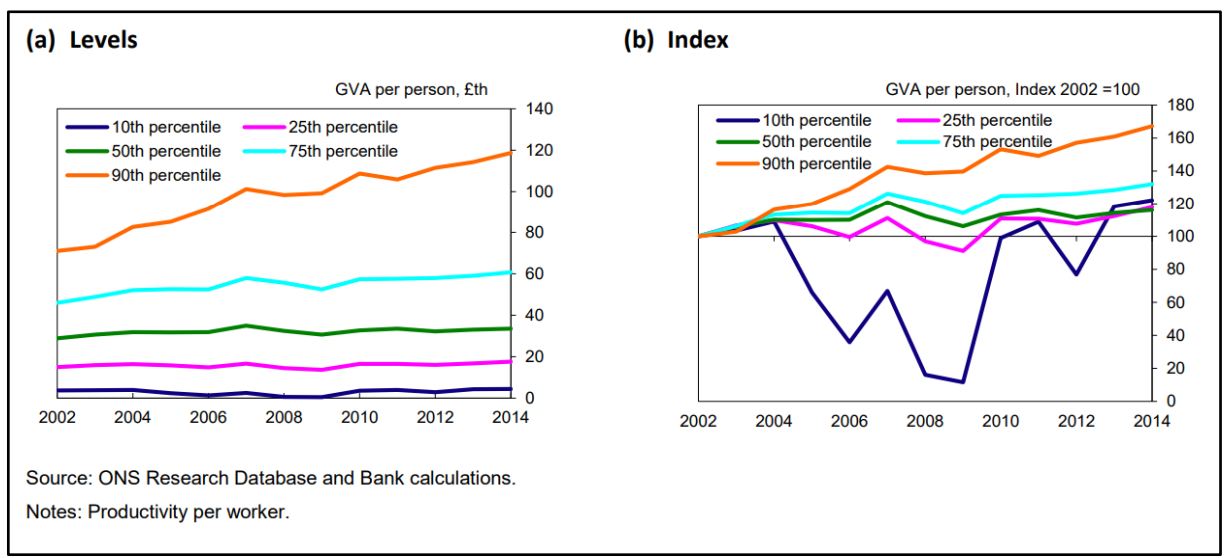

Given the link between technology and economic growth, a significant proportion of policy analysis is focussed on increasing technology adoption, innovation and diffusion throughout the economy. Policy makers require greater knowledge of the characteristics of these high-productivity firms and the conditions that support their success. An understanding of whether digital technologies, such as AI, are concentrated in highly productive firms would provide valuable evidence for policy design.

**Evidence of the impact of digital technology on productivity**

Evidence from the USA suggests that complementary investments in intangible digital assets, such as technical IT training and digital reorganisation, are significant drivers of productivity alongside tangible IT assets. Bjornolfson et.al show that the level of intangible

---

[9] Diagnosing the UK Productivity Slowdown: Which Sectors Matter and Why? Coyle, Mei (2022)
[10] Productivity puzzles. Speech given by Andrew G Haldane, Chief Economist, Bank of England (2017)
[11] The Best versus the Rest: Divergence across Firms during the Global Productivity Slowdown. Andrews, Criscuolo. CEP Discussion Paper No 1645 (2019)

assets (called digital capital) has grown considerably over the last 10 years. They also show that there is substantial heterogeneity in the levels of digital capital across firms, with much of it concentrated in the top decile of firms they call Superstar firms, akin to frontier firms mentioned previously. They find high a positive correlation between digital capital and firm-level productivity and strong predictive power of digital capital on productivity 3 years in the future, highlighting evidence of a time lag in the productive impact of these investments[12].

In the UK, Coyle et al. analysed whether digital-using firms are more productive[13]. Using data from the ONS Annual Business Survey, Annual Purchases Survey and e-commerce survey they study over 2,000 large UK firms between 2015 and 2018. Using two alternative measures of total factor productivity (TFP), and a bespoke measure of intangible capital stock, they assess the impact of a range of digital activities. They find large firms are more digitally intensive and more productive. They also find in-house digital capabilities are most important for driving productivity, which aligns with the findings of Bjornolfson that intangible digital capital is crucial. They also find evidence of a productivity j-curve for non-digitally intensive firms undertaking digital adoption, demonstrating a lag between adoption and productivity benefits being realised. Coyle's research finds a statistically significant positive relationship between adoption of digital technologies (e.g. cloud, e-commerce, big-data analytics) and TFP.

**Evidence of AI technology impact on productivity**

Alderucci et al. study the impact of AI innovating firms on productivity and labour demand. They use patent data to identify AI innovating firms and find that these firms "are on average much larger, older, better capitalised, more productive and significantly more likely to be multinational and multi-unit" than those without AI-related patents[14]. The research stops short of assigning causality between AI innovation and increased productivity due to data limitations, however their analysis shows likely indications of a positive relationship.

Using a similar methodology, analysis from Taiwan provides evidence of the positive causal relationship between AI innovation and productivity. Yang's research uses analytical techniques, such as generalised method of moments (GMM), to control for potential endogeneity in the results[15]. They find AI innovating firms are more productive, larger, pay higher salaries, and invest more in R&D than non-AI innovating counterparts. These findings corroborate those of Alderucci and other research such as Damiolo et al[16].

---

[12] Digital Capital and Superstar Firms. Tambe, Hitt, Rock, Brynjolfsson. NBER Working Paper No. 28285. (2020)
[13] Are digital-using UK firms more productive? Coyle (2022)
[14] Quantifying the Impact of AI on Productivity and Labor Demand: Evidence from U.S. Census Microdata. Alderucci et al. (2019)
[15] How Artificial Intelligence Technology Affects Productivity and Employment: Firm-level Evidence from Taiwan, Yang (2022)
[16] The impact of artificial intelligence on labor productivity, Damioli (2021)

Much of the current research identifies productivity impacts for those firms innovating in AI, often proxied via patents. Given the relatively small number of firms that are patenting AI technologies compared to the large number of firms that may adopt AI technology as a product or service, it is important to understand whether productivity benefits persist for this larger group of businesses.

Additionally, there is little evidence on how these firm-level impacts are translating into aggregate productivity impacts at the economy level. Brynjolfsson et.al. propose many possible reasons for this lack of aggregate impact, with the most compelling being time lags in causal effects and the required restructuring of the economy around new technologies[17]. This theory allows for increasing adoption of AI technology and some observable firm-level productivity impacts, without resulting impacts at an aggregate level due to the lack of intangible complementary investments that are required.

**How does AI impact productivity?**

Vital to assessing the impact of AI technologies is determining the mechanism by which AI impacts productivity. Corrado et al. consider AI as a mixture of tangible assets (e.g. hardware), measured intangible capital (e.g. software), and unmeasured intangible capital (e.g. databases)[18]. Brynjolfsson also argues that AI is an intangible asset, that should be accounted for in capital deepening but is often not fully measurable[19]. In many instances, measurable investments in tangible capital (e.g. robots, computers, servers) are accounted for as capital deepening, whilst unmeasured intangible inputs and outputs (e.g. improving representativeness of data, more efficient code) are not, therefore distorting productivity measurements. This is in line with the findings of Brynjolfsson (2020).

For example, AI technology being used in the transport and logistics industry for the optimisation of delivery routes can be thought of as capital deepening. Without investment in tangible capital (e.g. delivery vehicles) or an increase in labour inputs, the firm is able to increase output (e.g. deliveries). Investments in intangible capital (e.g. software) are the driver of the output increase, demonstrating an increase in labour productivity attributed to intangible capital deepening. However where the intangible inputs (e.g. data) and outputs (e.g. AI knowledge, patched software) are unmeasurable, this can impact productivity measurements.

The measurement of productivity is another challenge. Murray and Sharpe assess different productivity measures, highlighting that TFP is frequently associated with innovation and technological change[20]. Whilst this might make TFP an appropriate measure for productivity impacts, there are challenges in measurement and only growth

---

[17] Artificial Intelligence And The Modern Productivity Paradox: A Clash Of Expectations And Statistics. Brynjolfsson (2017)
[18] Artificial intelligence and productivity: an intangible assets approach. Corado, Haskell, Jona-Lasinio (2021)
[19] The Productivity J-Curve: How Intangibles Complement General Purpose Technologies. Brynjolfsson, Rock, Syverson (2020)
[20] Partial versus Total Factor Productivity Measures: An Assessment of their Strengths and weaknesses. Murray (2016)

rates can be determined, not absolute measures. Additionally it is hard to convey to non-experts due to its technical nature. Alternatively, factor-specific measures of productivity such as labour productivity are more intuitive (e.g. output per worker) and have practical implications for living standards, based on the prediction from the Cobb-Douglas production function that real wages will be proportional to labour productivity in the long-run.

Whilst the above discussion highlights the importance of disaggregating between TFP and capital deepening when measuring productivity, driven by data availability and In line with similar research in this area (Damioli et al.; Acemoglu and Restrepo[21]), this study uses labour productivity (measured in turnover per employee) as a productivity measure.

This literature shows converging evidence on the characteristics of firms adopting digital technologies and the positive relationship that digital technology adoption has on firm-level productivity. Whilst impacts on aggregate productivity measures are debated, valid theories provide rationale for why aggregate productivity improvements might not be observed. Whilst AI specific studies have been undertaken outside the UK, there is yet to be a firm-level analysis of the impact of AI adoption on UK labour productivity. This research conducts analysis to inform UK digital and technology policy on the links between AI adoption and productivity.

## Section 4: Research methodology

**Data**

A key barrier in conducting firm-level analysis on the impacts of AI is access to microdata for UK businesses. No database including data on AI usage, revenue, employment and other performance related figures exists. The complexities of linking multiple data sources are well documented and overcoming these provides significant value for economic researchers[22] [23].

This research links 2015-2019 data from the ONS e-commerce survey to the Inter-Departmental Business Register (IDBR)[24] [25]. The ONS e-commerce survey is UK-wide, covers all business sizes, and uses the IDBR as a sampling frame. A stratified sample is used, with stratas based on employment bands and SIC codes. Finance, Education, Healthcare, Agriculture and Arts & Entertainment industries are not covered by the survey.

The IDBR covers ~2.7 million UK businesses from all sectors of the economy. Businesses operating below VAT thresholds are not represented in the data and neither are non-profit organisations. The combined dataset is representative of UK businesses demographics in the sectors covered by the e-commerce survey.

---

[21] The Race between Man and Machine: Implications of Technology for Growth, Factor Shares, and Employment. Acemoglu, Restrepo (2018)
[22] Matching UK Business Microdata – A Study Using ONS and CBI Business Surveys. Mahony, Martin (2022)
[23] Artificial Intelligence, Labor, Productivity, and the Need for Firm- Level Data. Raj, Seamans (2019)
[24] E-commerce and ICT activity, UK Statistical bulletins, ONS
[25] Inter-Departmental Business Register (IDBR), ONS

Due to rotating questions in the e-commerce survey, this analysis uses two different samples of data to answer the research questions. Firstly it uses only 2019 data, covering 6,468 UK businesses, which includes questions specifically relating to the use of AI. Four questions relating to the use of AI techniques such as Machine Learning (ML), Natural Language Processing (NLP), 'other methods of big data analysis' and 'customer service chatbots' are used to construct a dummy variable for AI adoption. If a business responds yes to any of these questions they are considered to have adopted AI. This dataset is used to answer research question 1.

To answer research question 2, the analysis uses a balanced panel dataset from 2015-2019, including 457 UK businesses that produces 2,285 total observations. This data is used to assess whether those firms adopting AI technologies have higher labour productivity. Given the specific questions regarding AI are only included in the 2019 data, missing values are imputed backwards for each firm (e.g. If a firm uses AI in 2019, it is assumed they use it in all years from 2015-2019). Where there are missing values for some years for other variables, these are imputed forwards based on previous values for that business.

As a validation of the findings, a Data dummy variable has been used as a proxy for AI adoption. This is based on questions present in 2015 and 2019 relating to use of big data sources (whether a business collects smart-device data, geo-data, social media data or other big-data). The Data dummy has a correlation coefficient of 0.31 with the AI dummy variable and an R-Squared of 0.1 when regressing against the 2019 AI adoption dummy.

*Data limitations*

Given panel data is being used, the data may suffer from survivor bias and endogeneity in the attrition of observations and the dependent variable (labour productivity). Businesses that do not exist in all 5 years of the data are removed from the analysis, therefore the data may be biassed towards more successful businesses that may be included in each year's sample. However, much of the attrition is due to the sampling frame of the e-commerce survey being relatively small compared to the IDBR, therefore 5 years of consecutive responses for the e-commerce survey is rare.

Whilst efforts have been made to make the most of the data in relation to AI policy analysis, some of the analysis relies on assumptions that big-data and AI adoption are relatively interchangeable. Future analysis would benefit on questions specific to AI adoption running across multiple years and with more varied data types available e.g. numeric data rather than a dummy variable.

Given the e-commerce survey does not cover all sectors of the economy, the results are not representative of all UK businesses which must be considered when interpreting the results.

Finally, given survey data is being used rather than administrative data, the findings may be impacted from issues such as survey fatigue, misunderstanding of questions, or dishonest answers. The ONS is a trusted source for data so the expected impact of these limitations is small.

**Empirical strategy**

**Research question 1 - What are the key characteristics of firms adopting AI?**

The 2019 sample is used to undertake exploratory data analysis of businesses adopting AI. This aims to validate the findings in the literature review and provide understanding of key characteristics of firms using AI.

*Specification 1*

To analyse the relationship between firm characteristics and AI adoption, the specification includes key determinants found in the literature. Whilst the literature had a focus on AI innovating firms (e.g. those producing patents), this analysis looks at those adopting AI technologies.

Both models using in specification 2 are shown below, where $AI\_adoption_i$ is a dummy variable equal to 1 if a firm has adopted AI, $turnover_i$ is a numeric variable showing turnover, $employees_i$ is a numeric variable for the number of employees, and $Industry_i$ is a nominal categorical variable representing the Standard Industrial Classification (SIC) of an organisation.

A more comprehensive specification is estimated in model 2, including other firm-dependent variables as controls and dripping industry. The additional variables included in specification 2 are dummy variables, indicating whether a firm employs IT specialists ($EmployIT_i$), trains IT specialists ($TrainIT_i$), uses cloud computing ($Cloud_i$), analyses data from a variety of big-data sources ($Data_i$), or uses personalised content on their website ($Personalised_i$). These additional dummy variables attempt to include a measure of intangible digital capital in the specification.

*Specification 1, model 1:*

$$AI\_adoption_i = a + turnover_i + employees_i + Industry_i + u_i$$

*Specification 1, model 2:*

$$AI\_adoption_i = a + turnover_i + employees_i + EmployIT_i + TrainIT_i + Cloud_i + Data_i + Personalised_i + u_i$$

Given the dependent variable is binary, the estimation strategy uses a Linear Probability Model (LPM) measured with least squares. This model benefits from simplicity in its interpretation of coefficients:

> *"The change in percentage points of the probability of a "success" (i.e. y = 1) when the regressor, x, changes by one unit."*

Heteroscedasticity of the error terms is accounted for using robust standard errors. Any results will also be manually bounded by 0 and 1 given they represent probabilities. Robustness checks are carried out using Probit estimation, which is a nonlinear maximum-likelihood technique that can be used when dependent variables are binary. This was not chosen as the main estimation technique due to the complexity in interpretation of the findings.

**Research question 2 - What is the impact of adopting AI technologies on labour productivity?**

To estimate the impacts of AI adoption on labour productivity, the full panel dataset enables the analysis to control for both individual fixed effects and time fixed effects. Individual fixed effects are firm-level unobservable characteristics (e.g. management style, brand, culture) that are assumed to be constant over time but are correlated with the error term. Time fixed effects are factors that are constant across businesses but change over time (e.g. business cycles, price inflation).

Labour productivity is calculated as turnover divided by employment. A log transformation had been applied to normalise the distribution of the labour productivity variable. To test the robustness of the specifications, and to account for the time lag found in the literature, the specifications have been undertaken with a labour productivity measure that has no lag, a 1 year lag and a 2 year lag. This is due to uncertainty regarding the time taken for investment in AI technology to have a productivity impact, as discussed in section 3.

*Specification 2 - Pooled OLS estimator using AI dummy*

The effect of the AI dummy variable is first estimated using pooled Ordinary Least Squares (OLS). This groups all data points together and undertakes OLS estimation on the full panel dataset. This is undertaken using two different log-linear specifications, the first with only the AI dummy as an explanatory variable and the second including multiple control variables.

*Specification 2, Model 1:*

$$Log(Labour\ Productivity_{it}) = AI_i + u_{it}$$

*Specification 2, Model 2:*

$$Log(Labour\ Productivity_{it}) = AI_i + employees_{it} + Industry_{it} + Cloud_{it} + u_{it}$$

There are limitations of using pooled OLS on panel data. Firstly, because there are multiple observations from a single business this violates the assumption of independent errors, leading to the OLS standard errors to be incorrect. To account for correlation in standard errors, they are clustered at the individual business level. There is also a risk of omitted variables (observable or unobservable) causing the OLS estimators to be biassed and inconsistent.

Suitable estimation approaches are limited because the AI dummy is only available in 2019 data and has been imputed for all other time periods in the panel dataset, therefore is time-invariant and cannot be estimated using a Fixed Effects (FE) estimator. FE estimation cannot be undertaken because the effects of the AI dummy cannot be disentangled from the individual firm fixed effects and it is lost from the regression.

A Random Effects (RE) estimator is also not appropriate given endogeneity in the explanatory variables violates a key assumption in the use of Random Effects estimators. Endogeneity of some regressors was confirmed by implementing a Hausman Test, with results shown in figure 3. This shows the null hypothesis (zero correlation between the regressors and the error terms) is rejected, therefore suggesting that a FE model is more appropriate.

*Figure 3: Results of Hausman Test, Fixed Effects vs. Random Effects*

```
> phtest(ai_re_model, ai_fe_model,vcov. = vcovHC, type = "HC1", effect = "twoways" )

        Hausman Test

data:  labour_prod ~ AI + employee_band + cloud + training_it_special +  ...
chisq = 46.94, df = 12, p-value = 4.774e-06
alternative hypothesis: one model is inconsistent
```

Robustness checks are carried out by using pooled OLS whilst removing outlier observations with turnover above the 95th percentile. Removing these observations reduces the sample size considerably therefore these have been kept in the main analysis. There is further discussion on the robustness of results in section 5.

FE estimation techniques are used in specification 3, where the main explanatory variable of interest, the Data dummy, varies over time.

***Specification 3 - Within Group Fixed Effects using Data dummy***

Using the Data dummy variable, a FE model with Within Group (WG) estimation is undertaken. This is possible because there is variation in the Data dummy within a single business over the time-series, therefore the impact of the Data dummy can be disaggregated from the other time-invariant fixed effects of each business. The WG estimator approach de-means the data for each business, therefore removing the firm fixed effects. Given the within-firm fixed effects, standard errors are clustered at the firm level. Time fixed effects are also accounted for in the specification. First Difference (FD) estimation was used to validate the robustness of the findings. Similarly to specification 2, robustness checks are carried out by removing outliers.

Specification 3, model 1 only includes the Data dummy variable in the specification. Additional regressors, including employee band (Specification 3, model 2), adoption of Cloud computing, Training of IT specialists, industry and personalised content (Specification 3, model 3) are also introduced as controls.

***Specification 3, Model 1:***

$$Log(Labour\ Productivity_{it}) = Data_{it} + Firm\_fixed_i + Time\_fixed_t + u_{it}$$

***Specification 3, Model 2:***

$$Log(Labour\ Productivity_{it}) = Data_{it} + Firm\_fixed_i + Time\_fixed_t + emp\_band_{it} + u_{it}$$

***Specification 3, Model 3:***

$$Log(Labour\ Productivity_{it}) = Data_{it} + Firm\_fixed_i + Time\_fixed_t + emp\_band_{it} + Cloud_{it} + TrainIT_{it} + Industry_{it} + personalised_{it} + u_{it}$$

## Section 5: Research findings

**Descriptive analysis**

Figure 4 shows that businesses that have adopted AI, representing 11% of businesses in the sample, on average have turnover equal to ~£455 million and have ~1,750 employees. Both of these are considerably larger than firms that haven't adopted AI, which is consistent with the literature. This data also validates that AI is an emerging technology, adopted by a minority of the sample businesses (~11%).

*Figure 4: Summary statistics for 2019 data, AI adopters vs. non-AI adopters*

**Summary table of AI adopters vs. non-adopters**

| AI | count | Share (%) | Turnover (thousands) | Employment |
|---|---|---|---|---|
| FALSE | 5784 | 89 | 66661 | 318 |
| TRUE | 684 | 11 | 454767 | 1765 |

Figure 5 validates the positive relationship between AI adoption rates and firm size found in the literature. Only 3.5% of micro businesses have adopted AI in the UK, but this proportion increases steadily up to 21% in large firms

*Figure 5: Proportion of firms using AI, by size band*

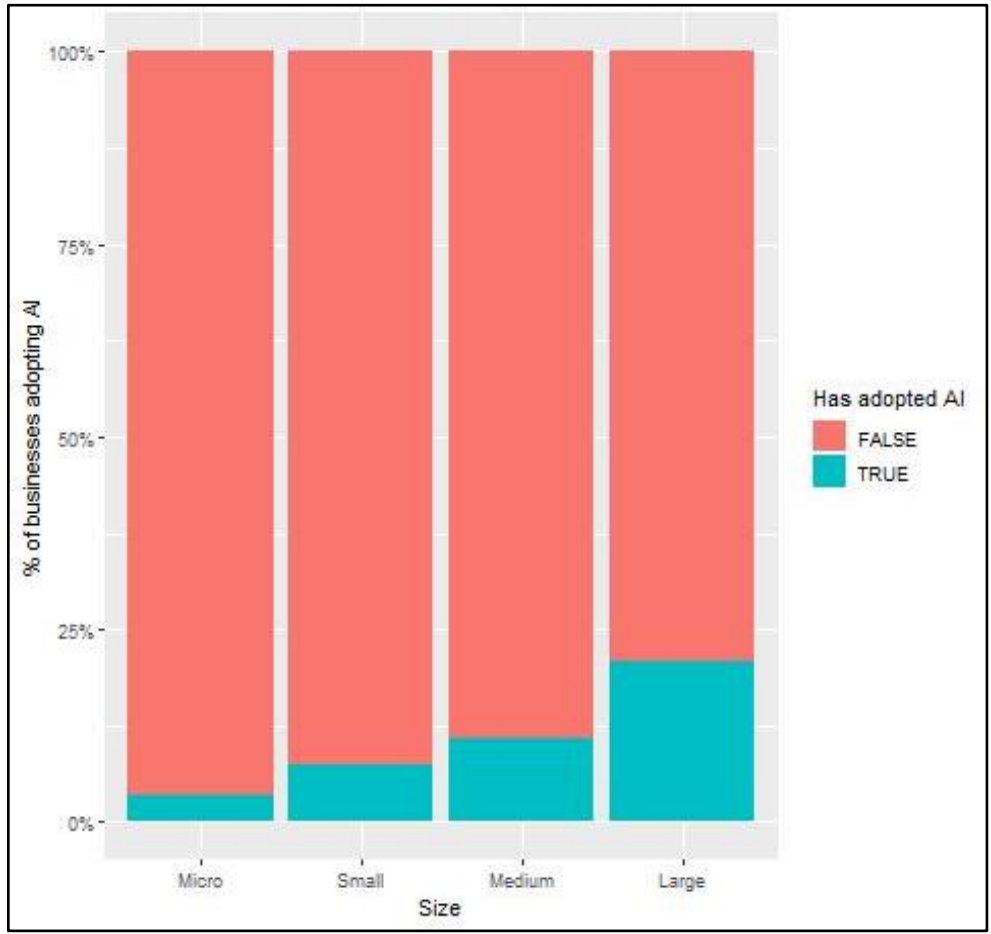

Figure 6 shows the proportion of businesses adopting AI by industry. Businesses in the Travel, IT, Trade and Administration industries have the highest adoption rates (>12%), whilst Construction, Manufacturing and Utilities have the lowest (<8%). This analysis shows that there are differences between adoption rates across industries, although the variability is lower than that found in other research. This could be due to the data not including Finance and Healthcare which are often found to have high adoption rates[26]. Also, given it is only a descriptive statistic, these changes could be driven by other factors such as industry demographic composition or further unobservable factors.

---

[26] AI Activity in UK Businesses, DCMS (2022)

*Figure 6: Proportion of firms using AI, by industry*

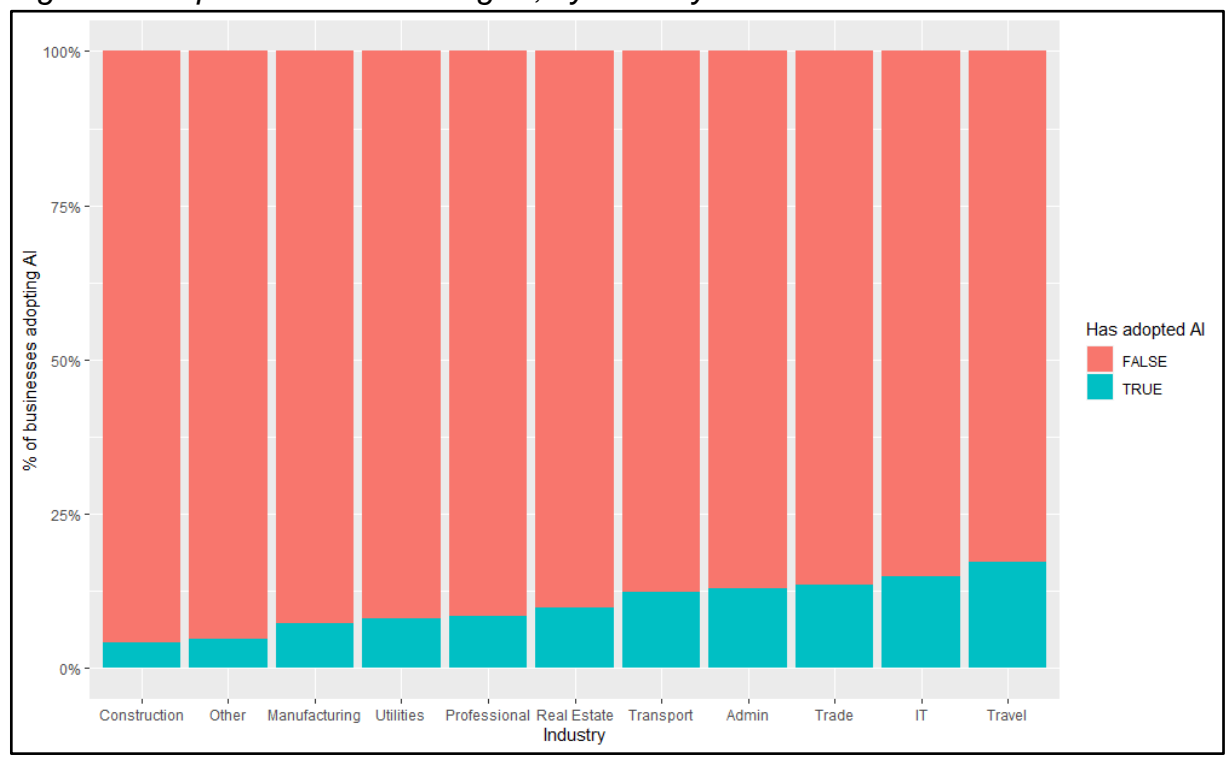

Of those firms that have adopted AI technologies, it is of interest to policy advisors to understand which AI techniques and applications they are using most. The most common approach was for AI to be implemented by an external business (66% of adopting businesses), followed by the use of chatbots (35%), other AI techniques (12%), Machine learning (10%) and NLP (6%)*. These findings validate recent DCMS research that identified ~40% of businesses developed in-house whilst the remaining ~60% outsourced or purchased "off-the-shelf" solutions from external businesses. These findings are interesting considering Coyle's (2019) work that identified businesses using external contractors experiencing a negative impact on productivity from digital adoption, and in-house capabilities being a critical driver of productivity gains from digital adoption.

This analysis applies a log transformation to the labour productivity variable given it has positive skewness (i.e. a long right tail). Figure 7 shows a histogram of log labour productivity for firms that have adopted AI to those that haven't in the left-hand chart, and those that have adopted big-data against those that haven't in the right-hand chart. It shows that firms that have adopted AI have a higher median log labour productivity than those that haven't adopted AI. This increase is consistent across both the 25th and 75th percentile, with the increase more pronounced at the 75th percentile. This same finding holds for big-data adopting businesses, although the difference is less pronounced.

Not shown in the chart, but in line with the finding of Coyle's 2021 paper, those businesses building AI internally have higher median labour productivity compared to all AI adopting businesses.

*Figure 7: Boxplot of log labour productivity, for AI and big-data adopting businesses*

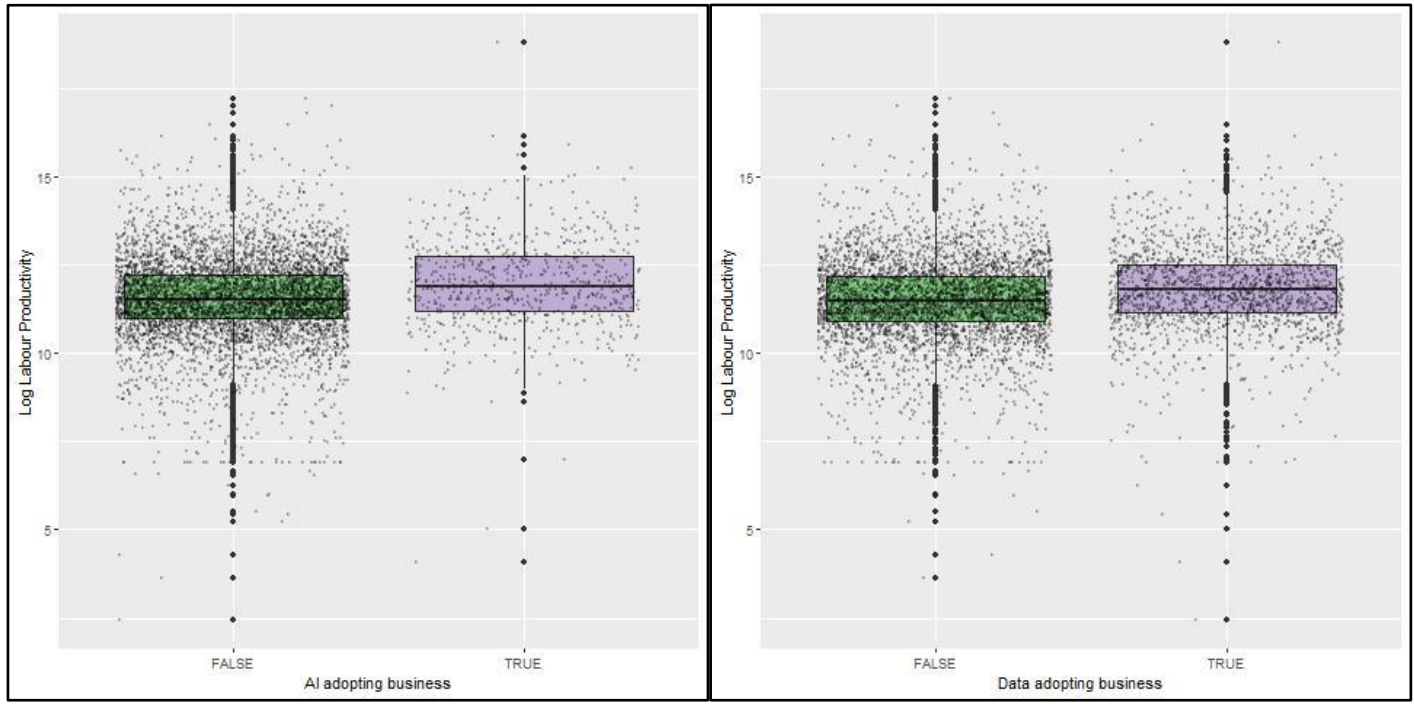

Figure 8 shows similar results, using the full panel dataset and displaying a density plot of log labour productivity for AI adopting firms (blue) and non-AI adopting firms (red). The mean of log labour productivity is also shown, highlighting a higher average log labour productivity in AI adopting businesses. Figure 8 also shows a larger proportion of firms with very high productivity have adopted AI, shown by a thicker right-hand tail. These businesses can be thought of as frontier or superstar firms discussed in the literature review.

*Figure 8: Density plot of log labour productivity, AI and non-AI firms (full panel dataset)*

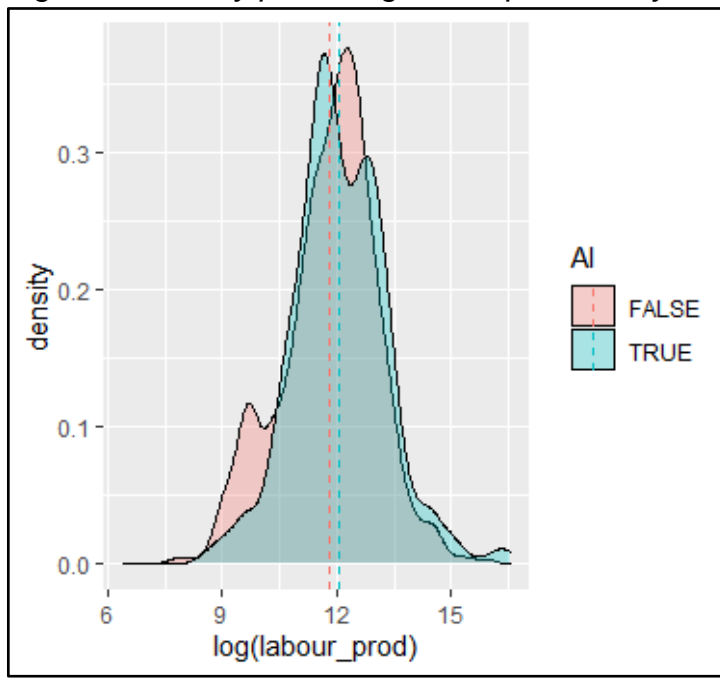

This descriptive analysis validates the findings in the literature that on average firms that have adopted AI technologies are on average bigger and more productive. However, it doesn't confirm any causal relationship between AI adoption and productivity. The relationships observed might be driven by omitted variables or simultaneity in the variables, whereby more productivity businesses may decide to adopt AI rather than AI adoption causing increased productivity. Econometric analysis will analyse these relationships with more robust techniques, to understand the explanatory strength of each independent variable.

**Research question 1 - What are the key characteristics of firms adopting AI?**

*Specification 1*

The findings in figure 9 from specification 1, model 1 show: 1) The variables included in model 1 explain ~4% of the variation in AI adoption between firms; 2) Both turnover and employment are positively correlated with AI adoption, both statistically significant at the 1% level; and, 3) Some industries have a statistically significant relationship with AI adoption, but others do not.

Whilst the number of regressors is relatively small, an R-squared of 0.04 is low. This shows the regressors only explain ~4% of the change in AI adoption. However, this is to be expected given decisions to adopt technology such as AI are complex, and are determined by multiple interrelated factors. Therefore whilst a small R-squared shows the model only has weak predictive power over whether a firm adopts AI, it does identify statistically significant variables.

Model 1 results show that both turnover and employment are statistically significant explanatory variables for AI adoption. When controlling for employment and industry, an increase in turnover of £1,000,000 is associated with an average increase in the probability of AI adoption by 0.01%. An increase of 1,000 employees is associated with an increase in the probability of AI adoption by 0.7%, when controlling for turnover and Industry. These are relatively small changes, but due to the large range of values these variables can take they still have significant predictive power.

The Industry dummy variable coefficients are interpreted in comparison to the admin industry, which is included in the intercept term to avoid the dummy variable trap. This analysis shows that Industry variability in AI adoption rates is not solely driven by the size composition of an industry. When controlling for turnover and employment, businesses in the construction industry are 8% less likely to have adopted AI than in the admin industry, whilst in manufacturing they are 5% less likely.

Specification 1, model 2 has greater explanatory power and provides further insights into the relationships between AI adoption and other digital capital proxy variables. Key findings from model 2 include: 1) 16% of the total variance in AI adoption can be explained by the model; 2) The use of big-data sources (Data dummy) has the highest positive correlation with AI adoption, statistically significant at the 1% level. If a business uses big-data, they are on average 17% more likely to have adopted AI than those that don't, when controlling for other variables. Both cloud and personalised content also have positive and statistically significant coefficients on AI adoption. 3) Employing IT specialists doesn't have a statistically significant impact on AI adoption, however providing training for IT specialists has a positive statistically significant impact at the 1% level. Businesses providing specialist IT training are on average 6% more likely to adopt AI than businesses that don't, once other factors are controlled for.

Whilst these findings describe in more detail the relationships between firm characteristics and AI adoption and establish empirical facts, they do not explain causal relationships between the variables.

*Figure 9: Specification 1, model 1 and 2 regression output table*

| Specification 1: Firm characteristics relationship with AI adoption | | |
|---|---|---|
| | \multicolumn{2}{c}{*Dependent variable:*} |
| | \multicolumn{2}{c}{AI} |
| | (1) | (2) |
| turnover_millions | 0.0001*** (0.00001) | 0.00003*** (0.00001) |
| employment_thousands | 0.007** (0.003) | 0.004** (0.002) |
| IndustryConstruction | -0.085*** (0.017) | |
| IndustryIT | 0.017 (0.021) | |
| IndustryManufacturing | -0.054*** (0.015) | |
| IndustryOther | -0.076** (0.036) | |
| IndustryProfessional | -0.039** (0.017) | |
| IndustryReal Estate | -0.027 (0.027) | |
| IndustryTrade | 0.005 (0.016) | |
| IndustryTransport | -0.012 (0.022) | |
| IndustryTravel | 0.042 (0.027) | |
| IndustryUtilities | -0.052** (0.022) | |
| it_specialists | | 0.013 (0.009) |
| training_it_special | | 0.064*** (0.013) |
| cloud | | 0.030*** (0.007) |
| data | | 0.172*** (0.010) |
| personalised_content | | 0.108*** (0.017) |
| Constant | 0.119*** (0.014) | -0.001 (0.004) |
| Observations | 6,468 | 6,468 |
| $R^2$ | 0.044 | 0.158 |
| Adjusted $R^2$ | 0.043 | 0.157 |
| Residual Std. Error | 0.301 (df = 6455) | 0.282 (df = 6460) |
| F Statistic | 24.985*** (df = 12; 6455) | 173.359*** (df = 7; 6460) |
| *Note:* | | *p<0.1; **p<0.05; ***p<0.01 |

**Research question 2 - What is the impact of adopting AI technologies on labour productivity?**

*Specification 2 - AI Dummy model*

Figure 10 shows specification 2, model 1 (represented by regression 1,3 and 5 in figure 10) and model 2 (represented by 2, 4 and 6) regression output for pooled OLS. The output table shows the results for all three log labour productivity measures (non-lagged, 1 year lag, 2 year lag).

In all regressions, the AI dummy has a positive coefficient on labour productivity statistically significant at the 5% level. This means the analysis finds we can reject the null hypothesis, that the impact of AI adoption on labour productivity is zero, with a 5% chance of a type 1 error (incorrectly rejecting the null hypothesis when it is actually correct). This indicates that AI adoption has a statistically significant relationship with labour productivity. The magnitude of the coefficient remains stable at 0.3 when AI adoption is a single regressor or when other control variables are included in the specification in model 2.

Due to the log-linear nature of the specification and AI being a dummy variable, the interpretation of the coefficient must be corrected using the Halvorsen-Palmquist correction technique (% change in Y from AI dummy being equal to 1 = 100* $\exp^{Beta}$ -1, where *Beta* is the coefficient in figure 10) [27]. Therefore, businesses that have adopted AI are expected to have 35% higher labour productivity compared to those that haven't adopted AI. This effect remains significant when controlling for the size of business, use of cloud computing and the industry which a business is in.

Other significant variables in the model suggest that differences in labour productivity are also affected by factors such as the size of business or their industry. These findings and the implications for policy are discussed further in section 6.

For reasons outlined in section 4, pooled OLS estimates can be biassed due to the violation of the assumption of strictly exogeneity. Biassed estimators imply the value of coefficients do not represent the true population values. Reasons for this endogeneity include unobservable omitted variables that influence both labour productivity and the decision to adopt AI technologies, selection into treatment or simultaneity in the regression.

Examples of omitted variables or selection into treatment, that are represented in the error term, include a firm's innovative culture or management practices. Because these characteristics are likely to be positively correlated with adopting AI technology and labour productivity, then the bias in the regression will overstate the impact of AI technology adoption on labour productivity. Simultaneity and reverse causality are discussed in more detail in the limitations section.

---

[27] The Interpretation of Dummy Variables in Semilogarithmic Equations, Halvorsen, Palmquist (1980)

More advanced econometric techniques, such as fixed effects estimation, are unable to identify the impact of AI adoption because this variable does not vary over time. This restricts the analysis that can be undertaken and introduces analytical limitations that are discussed later in the report.

*Figure 10: Specification 2, model 1 and 2, pooled OLS regression output*

|  | Dependent variable: | | | | | |
|---|---|---|---|---|---|---|
|  | log(labour_prod) | | log(labour_prod_lag1) | | log(labour_prod_lag2) | |
|  | (1) | (2) | (3) | (4) | (5) | (6) |
| AI | 0.3** (0.1) | 0.3** (0.1) | 0.3** (0.1) | 0.3** (0.1) | 0.3** (0.1) | 0.3** (0.1) |
| cloud |  | 0.2*** (0.1) |  | 0.2*** (0.1) |  | 0.2*** (0.1) |
| employee_bandMedium |  | 0.7** (0.3) |  | 0.6** (0.3) |  | 0.6* (0.3) |
| employee_bandSmall |  | -0.9* (0.5) |  | -0.6* (0.3) |  | -0.8*** (0.3) |
| IndustryConstruction |  | 1.9*** (0.3) |  | 1.9*** (0.3) |  | 1.9*** (0.3) |
| IndustryIT |  | 1.8*** (0.2) |  | 1.8*** (0.2) |  | 1.8*** (0.2) |
| IndustryManufacturing |  | 2.0*** (0.1) |  | 2.0*** (0.2) |  | 2.0*** (0.2) |
| IndustryOther |  | 1.8*** (0.5) |  | 1.8*** (0.5) |  | 1.8*** (0.5) |
| IndustryProfessional |  | 1.3*** (0.2) |  | 1.3*** (0.2) |  | 1.3*** (0.2) |
| IndustryReal Estate |  | 1.3*** (0.2) |  | 1.3*** (0.2) |  | 1.3*** (0.2) |
| IndustryTrade |  | 2.2*** (0.1) |  | 2.2*** (0.2) |  | 2.2*** (0.2) |
| IndustryTransport |  | 1.3*** (0.2) |  | 1.3*** (0.2) |  | 1.2*** (0.2) |
| IndustryTravel |  | 0.3* (0.2) |  | 0.3* (0.2) |  | 0.3* (0.2) |
| IndustryUtilities |  | 2.1*** (0.3) |  | 2.1*** (0.3) |  | 2.1*** (0.3) |
| Constant | 11.8*** (0.1) | 10.1*** (0.1) | 11.8*** (0.1) | 10.0*** (0.1) | 11.8*** (0.1) | 10.0*** (0.1) |
| Observations | 2,285 | 2,285 | 1,828 | 1,828 | 1,371 | 1,371 |
| R² | 0.01 | 0.5 | 0.01 | 0.4 | 0.01 | 0.4 |
| Adjusted R² | 0.01 | 0.5 | 0.01 | 0.4 | 0.01 | 0.4 |
| F Statistic | 21.8*** (df = 1; 2283) | 135.0*** (df = 14; 2270) | 17.0*** (df = 1; 1826) | 105.8*** (df = 14; 1813) | 12.2*** (df = 1; 1369) | 76.9*** (df = 14; 1356) |

Note: *p<0.1; **p<0.05; ***p<0.01

***Specification 3 - Within Group Fixed Effects Data Dummy model***

Figure 11 displays regression summaries for the three models in specification 3. Results are shown for all labour productivity variables (no lag, 1 year lag and 2 year lag). Specification 3 models include:

- *Specification 3, Model 1 (represented by 1,4 and 7 in figure 11)* - Only includes Data dummy variable.

- *Specification 3, Model 2 (represented by 2,5 and 8 in figure 11)* - Also Includes firm size as regressor.

- *Specification 3, Model 3 (represented by 3,6 and 9* in figure 11) - Also Includes firm size, industry, cloud, personalised content and training for IT specialists as regressors.

Across all three models, and all variations in the lag of the dependent variable, the coefficient on the Data dummy remains statistically insignificant.

When using either non-lagged or 1 year lag on the labour productivity variable, the coefficient on the Data dummy is negative but not statistically significant across all three models (1-6 in figure 11). When using a 2 year lag on labour productivity, the coefficient on the Data dummy turns positive, but remains statistically insignificant across all three models. This indicates that when accounting for individual-firm fixed effects and time fixed effects, the use of big-data does not impact labour productivity.

Some of the control variables have a statistically significant relationship with labour productivity. The IT industry dummy is statistically significant at the 1 % level using all measures of labour productivity, but using non-lagged labour productivity the coefficient is negative, however when using a 1 and 2 year lag the coefficient becomes increasingly positive. This could indicate increasing productivity in the IT sector over the sample period.

Specification 3 has shown that the Data dummy variable does not have a statistically significant impact on labour productivity under a range of different model specifications. This implies that the use of big-data does not have a significant impact on labour productivity, once the individual firm and time fixed effects have been accounted for. These findings and the implications for policy are discussed further in section 6.

*Figure 11: Specification 3 regression output table*

| | Dependent variable: | | | | | | | | |
|---|---|---|---|---|---|---|---|---|---|
| | log(labour_prod) | | | log(labour_prod_lag1) | | | log(labour_prod_lag2) | | |
| | (1) | (2) | (3) | (4) | (5) | (6) | (7) | (8) | (9) |
| data | -0.01 | -0.01 | -0.01 | -0.01 | -0.01 | -0.01 | 0.0003 | 0.0004 | 0.003 |
| | (0.02) | (0.02) | (0.02) | (0.02) | (0.02) | (0.02) | (0.03) | (0.03) | (0.03) |
| employee_bandMedium | | 0.1 | 0.1 | | 0.02 | 0.03 | | -0.02 | -0.02 |
| | | (0.1) | (0.1) | | (0.03) | (0.03) | | (0.02) | (0.03) |
| employee_bandSmall | | -0.4 | -0.4 | | $0.1^{***}$ | $0.1^{***}$ | | | |
| | | (0.4) | (0.4) | | (0.04) | (0.04) | | | |
| cloud | | | 0.002 | | | -0.03 | | | -0.04 |
| | | | (0.02) | | | (0.02) | | | (0.02) |
| training_it_special | | | 0.003 | | | 0.02 | | | -0.02 |
| | | | (0.03) | | | (0.04) | | | (0.02) |
| IndustryConstruction | | | -0.3 | | | | | | |
| | | | (0.3) | | | | | | |
| IndustryIT | | | $-0.3^{***}$ | | | $0.2^{***}$ | | | $0.5^{***}$ |
| | | | (0.02) | | | (0.02) | | | (0.02) |
| IndustryManufacturing | | | $0.3^{*}$ | | | $0.2^{**}$ | | | -0.1 |
| | | | (0.1) | | | (0.1) | | | (0.2) |
| IndustryProfessional | | | 0.1 | | | $-0.1^{***}$ | | | -0.2 |
| | | | (0.1) | | | (0.02) | | | (0.2) |
| IndustryTrade | | | $0.5^{***}$ | | | 0.6 | | | 0.3 |
| | | | (0.2) | | | (0.4) | | | (0.3) |
| IndustryTransport | | | 0.03 | | | -0.1 | | | -0.3 |
| | | | (0.3) | | | (0.5) | | | (0.3) |
| IndustryTravel | | | 0.1 | | | -0.03 | | | 0.4 |
| | | | (0.1) | | | (0.03) | | | (0.3) |
| personalised_content | | | 0.01 | | | 0.01 | | | 0.03 |
| | | | (0.03) | | | (0.03) | | | (0.1) |
| Observations | 2,285 | 2,285 | 2,285 | 1,828 | 1,828 | 1,828 | 1,371 | 1,371 | 1,371 |
| $R^2$ | 0.0001 | 0.004 | 0.01 | 0.0001 | 0.0002 | 0.01 | 0.0000 | 0.0000 | 0.02 |
| Adjusted $R^2$ | -0.3 | -0.2 | -0.2 | -0.3 | -0.3 | -0.3 | -0.5 | -0.5 | -0.5 |
| F Statistic | 0.1 (df = 1; 1823) | $2.3^{*}$ (df = 3; 1821) | $1.9^{**}$ (df = 13; 1811) | 0.1 (df = 1; 1367) | 0.1 (df = 3; 1365) | $1.6^{*}$ (df = 12; 1356) | 0.0002 (df = 1; 911) | 0.005 (df = 2; 910) | $1.9^{**}$ (df = 11; 901) |

Note: $^{*}$p<0.1; $^{**}$p<0.05; $^{***}$p<0.01

**Limitations and robustness of results**

Some limitations remain in the analysis that need to be considered when interpreting the results. The labour productivity measure used (turnover / number of employees) is a simplification and, for example, does not account for hours worked per employee. This limitation exists in the data and therefore cannot be corrected for in the analysis. If hours worked per employee are endogenous to the regression then this may influence the findings.

Complimentary business investments (e.g. capital deepening) are not accounted for in the analysis given this data was not available in the data. The omission of this data may mean incorrect inferences are made about the impact of included variables.

Simultaneous causality may exist in the regression specifications, especially 2 and 3. For example, more productive firms may have the ability to adopt AI, *and* AI may increase productivity. Additionally, selection into treatment may occur. This is because the decision to adopt AI is not a random process and may be correlated with highly motivated and innovative businesses, which are also correlated with more productive firms. These issues can introduce endogeneity into the regression equation which can bias estimates.

The findings of this analysis do not prove causal relationships between variables and further analysis is required to improve on these results. Limitations in the AI variable only being available in a single year of the panel and a lack of numeric data on AI adoption (e.g. spend on AI technology) restricted the findings from the analysis, specifically not being able to undertake FE estimation on the AI adoption dummy variable.

Specification 1 has also been estimated using Probit estimation (Figure A1 in Annex 1) which validates the findings that firms with higher turnover and employment are more likely to adopt AI. It also validated the positive relationship between the use of digital variables (e.g.training IT specialists, cloud and big-data) and AI adoption.

Robustness checks were undertaken by estimating Specification 2, when removing observations with turnover bigger than the 95th percentile (Figure A2 in Annex 1). This shows that the AI dummy is still positive and statistically significant at the 10% level when additional controls are introduced in model 2, but loses the statistically significant effect in model 1 as a single regressor. Due to the nature of the balanced panel data, removing these outliers substantially reduces the sample size (from 457 to 293 businesses) which may impact results. This change in statistical significance could be due to removing the 'frontier firms' from the sample, which the literature review showed are more likely to use AI.

All three models of specification 3 were also estimated using a First Difference (FD) estimator. The results of FD estimation validated the findings of the WG estimation, with the Data dummy not having a statistically significant effect on labour productivity. The results of this estimation are included in annex 1 (figure A3), with only the non-lagged estimation included due to their similarity to the WG results. This increases the robustness of the conclusions that big-data is not a significant determinant of labour productivity.

## Section 6: Project outcomes

This project aimed to answer two research questions
   1. What are the key characteristics of firms adopting AI?
   2. What is the impact of adopting AI technologies on labour productivity?

**What are the key characteristics of firms adopting AI?**

Specification 1 identified that larger firms were more likely to have adopted AI. This analysis found positive and statistically significant correlations of AI adoption and both employee numbers and revenue. The analysis also identified that businesses in certain industries were more likely to have adopted AI.

Adoption of other digital technologies including cloud, big data and personalised content, and the training of IT specialists were also positively correlated with AI adoption. These variables can be viewed as contributing to digital capital, therefore validating the findings in the literature.

These findings indicate an opportunity for government policy to support smaller businesses to adopt AI and suggest that they might face barriers that larger businesses do not. The findings also suggest that there is an opportunity for targeted government support in industries that have lower AI adoption levels such as construction, manufacturing or utilities.

Whilst these results don't prove causal relationships due to the limitations outlined, they validate many of the findings in the literature and provide greater confidence that these findings apply to UK businesses adopting AI.

**What is the impact of adopting AI technologies on labour productivity?**

Specification 2 showed that the AI dummy variable was a statistically significant predictor of labour productivity. The analysis estimated that businesses using AI had, on average, 35% higher labour productivity than those that didn't use AI. These findings were robust to a variety of different specifications and lagged dependent variables. This finding validates many of the relationships shown in the descriptive analysis section and discussed in the literature. However, data limitations restrict the use of more robust statistical techniques, such as within-group or first-difference estimation to control for individual firm fixed-effects, to more definitively confirm this relationship or identify causal effects.

Care should be taken in generalising this result. This analysis didn't take into account costs of production, such as cost of labour and capital, which may have important implications for the conclusions. Also, the productivity measure used in this analysis represents revenue per employee, rather than a more sophisticated measure of productivity.

Specification 3 found that the use of big-data did not have a statistically significant effect on labour productivity under a range of different model specifications. Once the analysis accounted for individual business fixed effects and time fixed effects, big-data was not a determinant of labour productivity. Neither were the use of cloud computing or personalised content.

The change in interpretation of the findings for the cloud computing variable between specification 2 and 3 shows that care must be taken when assigning confidence to the results for AI adoption. Whilst the analysis suggests a significant effect of cloud computing in specification 2, this effect disappears in specification 3 once firm fixed effects are accounted for. This may also be the case for AI adoption, although data limitations prevent this from being confirmed.

These findings suggest that the relationship between digital technologies and productivity is complex. Whilst encouraging the use of big-data sources may not be enough to increase productivity by itself, the use of AI technologies could have a large impact. A greater focus should be placed on understanding the determinants of labour productivity at an individual business level, as differences may be caused by other important factors.

These results also highlight the importance of robust quantitative analysis when determining impacts and interpreting relationships in data. It has shown that incorrect inferences can be made if thorough analysis is not conducted and contributing factors are not accounted for. Due to the discrepancies between the findings between specification 2 and 3, subsequent analysis should look to analyse the impact of AI adoption using more sophisticated techniques that control for firm fixed effects. This would provide greater certainty in the results.

**Achievement of objectives**

Whilst this research set out to answer the above research questions, it also aimed to achieve a set of objectives to provide tangible benefits. These objectives were:

1. Develop theoretical framework for AI adoption and business decision making
2. Create a unique panel micro-dataset for UK businesses, including AI adoption and key business variables
3. Establish and estimate an econometric specification to answer the research questions
4. Interpret and consider the key policy implications of the findings

This research succeeded in establishing a theoretical framework for businesses decision making regarding AI adoption. It used empirical evidence from the literature to validate this theoretical framework and ensure that the research specifications were backed by findings from relevant literature.

The research also succeeded in creating a unique panel micro-dataset using multiple ONS surveys to link digital technology adoption to labour productivity variables. This dataset will provide ongoing value for future analysis within the Office for AI and Digital and Technology Policy directorate in government.

The research used advanced statistical knowledge such as Linear Probability Models, OLS, pooled OLS, Within Group estimation, Probit estimation, and robust and clustered standard errors. These advanced techniques ensure correct interpretation of results, and provide greater insight into the relationship between variables by controlling for individual-firm fixed effects and time-fixed effects in the results. This enables analysis to go beyond simple visualisations and correlations to provide greater confidence in causal relationships. Whilst the results of the analysis weren't conclusive, they provide important progress in understanding the determinants of AI adoption and its effect on productivity.

Finally, these results of this analysis have been put into context through application to the policy questions this research set out to answer. This analysis has increased the level of insight that policy officials have on the determinants of AI adoption and the potential impacts of AI technology on labour productivity. These findings will be used to influence the design and implementation of policy measures to help the Office for AI achieve its objectives.

## Section 7: Recommendations and conclusions

**Recommendations**

A set of recommendations have been identified as a result of undertaking this analysis. These have been grouped by the relevant stakeholders to consider.

**Digital analysis team**

The research has shown that data limitations can create barriers to effective analysis. It has also proven that existing data sources from the ONS can provide valuable insights if sufficient data processing and linking is undertaken. Ultimately, it has shown that using more advanced analytical techniques enables greater insights from data and allows greater confidence in the findings from quantitative analysis. Specific recommendations that should be taken forwards by the digital analysis team include:

1. Improve the availability of reliable data on digital technology adoption.
2. Better utilise ONS microdata.
3. Develop greater econometric skills throughout the analytical team.
4. Use new data releases to validate these research findings in subsequent years.

**The Office for AI (OAI) policy team**

Whilst data analysis and quantitative findings can provide insights into the relationships of variables, these need to be supplemented with qualitative understanding in order to form well evidenced government policy. The following recommendations should be considered by the OAI policy team:

1. Discuss the barriers to AI adoption with small and medium sized businesses.
2. Determine the reasons that some industries are lagging behind in AI adoption and consider policy interventions that could increase adoption levels.

**Conclusions**

The adoption of digital technologies is expected to continue at a rapid rate. Whilst greater use of innovative technologies might be expected to bring productivity benefits based on macroeconomic and microeconomic theory, the results from empirical evidence are less conclusive.

The results from this study demonstrate that the adoption of AI technologies is concentrated in larger businesses and in certain industries. It also found that the activities building up digital capital (e.g. adoption of digital technology such as cloud computing or big-data, or training IT specialists) were positively associated with the adoption of AI technology. These findings validate the literature and provide clear insight to help target policies aimed at increasing the adoption of AI technology.

AI adoption was found to have a significant positive effect on labour productivity, whereas the use of big-data was not. Limitations in the availability of data, and the restrictions this had on the analytical techniques used, means there is still uncertainty on the robustness of the findings for AI adoption. Concerted effort to improve the frequency of AI related data and the type of data available could bring improvements in future analytical research.

Whilst this research has been quantitative in nature, additional qualitative analysis should be undertaken to provide reasoning and rationale for the findings. Qualitative insights that validate the quantitative results would improve the robustness of the results.

This research has shown that increasing the adoption of AI technologies may help to resolve the UK's productivity problems, but is not expected to be a silver bullet. There are multiple factors that influence productivity, and whilst AI technologies might be part of that, the analysis has shown that this relationship is difficult to isolate and quantitatively prove.

The findings of this research will build on the understanding that policy teams have and contribute to improving the evidence base for digital and technology policy. The objectives achieved as part of this study will support subsequent analysis into the determinants of productivity and whether digital technologies have a significant effect on this. The recommendations outlined should be considered to ensure that the findings from this research are incorporated into policy design and future analysis.

# Annex 1: Additional figures

*Figure A1: Specification 1 - Probit estimation*

**Specification 1 - Probit estimation: Firm characteristics relationship with AI adoption**

| | Dependent variable: AI | |
|---|---|---|
| | (1) | (2) |
| turnover_millions | 0.0003** (0.0001) | 0.0001* (0.00005) |
| employment_thousands | 0.081*** (0.027) | 0.035*** (0.013) |
| IndustryConstruction | -0.587*** (0.130) | |
| IndustryIT | 0.107 (0.099) | |
| IndustryManufacturing | -0.295*** (0.085) | |
| IndustryOther | -0.474 (0.344) | |
| IndustryProfessional | -0.194** (0.092) | |
| IndustryReal Estate | -0.124 (0.153) | |
| IndustryTrade | 0.042 (0.081) | |
| IndustryTransport | -0.054 (0.113) | |
| IndustryTravel | 0.158 (0.114) | |
| IndustryUtilities | -0.302** (0.141) | |
| it_specialists | | 0.186*** (0.067) |
| training_it_special | | 0.243*** (0.067) |
| cloud | | 0.239*** (0.055) |
| data | | 0.894*** (0.049) |
| personalised_content | | 0.424*** (0.063) |
| Constant | -1.230*** (0.071) | -2.097*** (0.049) |
| Observations | 6,468 | 6,468 |
| Log Likelihood | -2,055.785 | -1,737.932 |
| Akaike Inf. Crit. | 4,137.569 | 3,491.863 |

*Note:* *p<0.1; **p<0.05; ***p<0.01

*Figure A2: Specification 2, Pooled OLS with observations with turnover > 95th percentile removed*

|  | Dependent variable: | | | | | |
|---|---|---|---|---|---|---|
|  | log(labour_prod) | | log(labour_prod_lag1) | | log(labour_prod_lag2) | |
|  | (1) | (2) | (3) | (4) | (5) | (6) |
| AI | 0.1 | 0.2* | 0.1 | 0.2* | 0.1 | 0.2* |
|  | (0.2) | (0.1) | (0.2) | (0.1) | (0.2) | (0.1) |
| cloud |  | 0.2*** |  | 0.2*** |  | 0.2*** |
|  |  | (0.1) |  | (0.1) |  | (0.1) |
| employee_bandMedium |  | 1.0*** |  | 1.0*** |  | 1.0*** |
|  |  | (0.3) |  | (0.3) |  | (0.3) |
| employee_bandSmall |  | -0.3 |  | 0.1 |  | -0.03 |
|  |  | (0.4) |  | (0.3) |  | (0.3) |
| IndustryConstruction |  | 1.3*** |  | 1.3*** |  | 1.3*** |
|  |  | (0.2) |  | (0.2) |  | (0.2) |
| IndustryIT |  | 1.4*** |  | 1.4*** |  | 1.5*** |
|  |  | (0.2) |  | (0.2) |  | (0.2) |
| IndustryManufacturing |  | 1.9*** |  | 1.9*** |  | 1.9*** |
|  |  | (0.1) |  | (0.1) |  | (0.1) |
| IndustryOther |  | 1.1*** |  | 1.2*** |  | 1.2*** |
|  |  | (0.1) |  | (0.1) |  | (0.1) |
| IndustryProfessional |  | 1.0*** |  | 1.0*** |  | 1.0*** |
|  |  | (0.2) |  | (0.2) |  | (0.2) |
| IndustryReal Estate |  | 1.3*** |  | 1.3*** |  | 1.3*** |
|  |  | (0.2) |  | (0.2) |  | (0.2) |
| IndustryTrade |  | 2.1*** |  | 2.1*** |  | 2.1*** |
|  |  | (0.1) |  | (0.1) |  | (0.2) |
| IndustryTransport |  | 1.1*** |  | 1.1*** |  | 1.1*** |
|  |  | (0.2) |  | (0.2) |  | (0.2) |
| IndustryTravel |  | 0.4** |  | 0.4** |  | 0.4** |
|  |  | (0.2) |  | (0.2) |  | (0.2) |
| IndustryUtilities |  | 1.5*** |  | 1.4*** |  | 1.4*** |
|  |  | (0.3) |  | (0.4) |  | (0.4) |
| Constant | 11.4*** | 10.0*** | 11.4*** | 9.9*** | 11.4*** | 9.9*** |
|  | (0.1) | (0.1) | (0.1) | (0.1) | (0.1) | (0.1) |
| Observations | 1,465 | 1,465 | 1,172 | 1,172 | 879 | 879 |
| $R^2$ | 0.001 | 0.5 | 0.001 | 0.5 | 0.001 | 0.5 |
| Adjusted $R^2$ | 0.0002 | 0.5 | 0.0001 | 0.5 | -0.0001 | 0.5 |
| F Statistic | 1.3 (df = 1; 1463) | 120.9*** (df = 14; 1450) | 1.2 (df = 1; 1170) | 94.0*** (df = 14; 1157) | 0.9 (df = 1; 877) | 67.8*** (df = 14; 864) |

Note: *p<0.1; **p<0.05; ***p<0.01

*Figure A3: Specification 3, First Difference regression output table (non-lagged labour productivity)*

|  | Dependent variable: | | |
|---|---|---|---|
|  | log(labour_prod) | | |
|  | (1) | (2) | (3) |
| data | 0.002 | 0.002 | 0.003 |
|  | (0.02) | (0.02) | (0.02) |
| employee_bandMedium |  | 0.1* | 0.1* |
|  |  | (0.1) | (0.1) |
| employee_bandSmall |  | -0.1 | -0.1 |
|  |  | (0.2) | (0.2) |
| cloud |  |  | 0.04** |
|  |  |  | (0.02) |
| training_it_special |  |  | -0.002 |
|  |  |  | (0.02) |
| IndustryConstruction |  |  | -0.5*** |
|  |  |  | (0.1) |
| IndustryIT |  |  | -0.3*** |
|  |  |  | (0.01) |
| IndustryManufacturing |  |  | 0.1 |
|  |  |  | (0.1) |
| IndustryProfessional |  |  | 0.1 |
|  |  |  | (0.1) |
| IndustryTrade |  |  | 0.1 |
|  |  |  | (0.1) |
| IndustryTransport |  |  | -0.1 |
|  |  |  | (0.1) |
| IndustryTravel |  |  | 0.02 |
|  |  |  | (0.03) |
| personalised_content |  |  | 0.001 |
|  |  |  | (0.02) |
| Constant | 0.04*** | 0.04*** | 0.04*** |
|  | (0.01) | (0.01) | (0.01) |
| Observations | 1,828 | 1,828 | 1,828 |
| $R^2$ | 0.0000 | 0.001 | 0.01 |
| Adjusted $R^2$ | -0.001 | -0.0003 | -0.0002 |
| F Statistic | 0.01 (df = 1; 1826) | 0.8 (df = 3; 1824) | 1.0 (df = 13; 1814) |

Note: *p<0.1; **p<0.05; ***p<0.01

## Annex 2: References